\documentclass[fleqn,usenatbib]{mnras}

\usepackage{newtxtext,newtxmath}

\usepackage[T1]{fontenc}

\DeclareRobustCommand{\VAN}[3]{#2}
\let\VANthebibliography\thebibliography
\def\thebibliography{\DeclareRobustCommand{\VAN}[3]{##3}\VANthebibliography}

\usepackage{graphicx}	
\usepackage{amsmath}	
\usepackage[version=4]{mhchem}
\usepackage{color}

\title[Solenoidal modes and star formation efficiency]{Solenoidal turbulent modes and star formation efficiency in Galactic-plane molecular clouds}

\author[R. Rani et al.]{
Raffaele Rani,$^{1,2}$\thanks{E-mail: rani@ntnu.edu.tw (RR)}
Toby J. T. Moore,$^{2}$
David J. Eden, $^{2,3}$
Andrew J. Rigby $^{4}$
\\
$^{1}$Center of Astronomy and Gravitation, Department of Earth Sciences, National Taiwan Normal University, 88, Sec. 4, Ting-Chou Rd., Wenshan District, \\ Taipei 116, Taiwan R.O.C.\\
$^{2}$Astrophysics Research Institute, Liverpool John Moores University, IC2, Liverpool Science Park, 146 Brownlow Hill, Liverpool, L3 5RF, U.K.\\
$^{3}$Armagh Observatory and Planetarium, College Hill, Armagh, BT61 9DB, UK\\
$^{4}$ School of Physics and Astronomy, Cardiff University, Queen’s Building, The Parade, Cardiff, CF24 3AA, UK \\}

\date{Accepted XXX. Received YYY; in original form ZZZ}

\pubyear{2022}

\begin{document}

\label{firstpage}
\pagerange{\pageref{firstpage}--\pageref{lastpage}}
\maketitle

\begin{abstract}

It is speculated that the high star-formation efficiency observed in spiral-arm molecular clouds is linked to the prevalence of compressive (curl-free) turbulent modes, while the shear-driven solenoidal (divergence-free) modes appear to be the main cause of the low star-formation efficiency that
characterises clouds in the Central Molecular Zone. Similarly, analysis of the Orion B molecular cloud has confirmed that, although turbulent modes vary locally and at different scales within the cloud, the dominant solenoidal turbulence is compatible with its low star formation rate. This evidence points to inter-and intra-cloud fluctuations of the solenoidal modes being an agent for the variability of star formation efficiency. We present a quantitative estimation of the relative fractions of momentum density in the solenoidal modes of turbulence in a large sample of plane molecular clouds in the \ce{^{13}CO}/\ce{C^{18}O}
($J=3\rightarrow 2$) Heterodyne Inner Milky Way Plane Survey 
(CHIMPS). We find a negative correlation between the solenoidal fraction and star-formation efficiency. This feature is consistent with the hypothesis that solenoidal modes prevent or slow down the collapse of dense cores. In addition, the relative power in the solenoidal modes of turbulence (solenoidal fraction) appears to be higher in the Inner Galaxy declining with a shallow gradient with increasing Galactocentric distance. Outside the Inner Galaxy, the slowly, monotonically declining values suggest that the solenoidal fraction is unaffected by the spiral arms.

\end{abstract}




\section{Introduction}
The conversion of molecular gas into stars is one of the fundamental baryonic processes that shape the visible Universe, driving cosmic evolution from the epoch of re-ionisation to present-day Galactic systems. The earliest stages of star formation see neutral gas in the interstellar medium (ISM) aggregating in dense molecular clouds through large-scale hydrodynamic, thermodynamic, or gravitational instabilities. These perturbations are associated with colliding, or shearing flows or shocks caused by the gas entering the spiral arms. Dissipative shocks in the supersonic turbulence resulting from the cloud-formation process,  then \cite[or concurrently,][]{Heitch2008} form fragmented, compressed layers, and filaments. Dense fragments become gravitationally self-bound and collapse into the clumps and cores that eventually create stars, while more rarefied structures are transient and dissipate.  As the view of molecular clouds as naturally largely transient features has succeeded their older characterisation as extant structures in a state of quasi-equilibrium preceding collapse, it has become clear they are supported by the interplay of factors acting on different scales. 

Molecular clouds have highly irregular and complex shapes. Many of them possess wispy filamentary structures that resemble those of atmospheric clouds. The irregular boundaries of molecular clouds found on contour maps show fractal properties \citep{Dickman1990, Falgarone1991, Falgarone1992, Zimmermann1992, Elia2018}. The fractal dimension estimated for clouds has similar values to those found at various interfaces in
turbulent flows \citep{Falgarone1991, Sreenivasan1991}, 
suggesting that turbulence plays a fundamental role in the formation and evolution of molecular clouds. 

Commonly, velocity dispersions within molecular clouds are about ten times larger than expected by solely considering thermal properties \citep{Larson1981, Rathborne2009}. This is generally interpreted as evidence of turbulence being a prominent factor in creating and sustaining a cloud's internal structure. In this picture, the evolution of molecular clouds and the star-forming regions within them is governed by the complicated interactions of gravity, magnetic fields \citep{Elmegreen2004, MacLow2004, McKee2007, Heyer2012} and supersonic turbulent motions driven at different scales from stellar feedback to Galactic shear \citep{Scalo2004}. On microscopic scales, the interactions between the gas molecules with the surrounding far-UV and cosmic-ray radiation regulate the thermodynamic state of the gas and its coupling to magnetic fields. 

Despite the progress in the characterisation of molecular clouds and their structure, devising a quantitative model, empirical or theoretical, that predicts the efficiency of star-forming processes and their relation to the physical properties of the interstellar gas is an elusive task. Empirical relations such as Schmidt-Kennicutt \citep{Kennicutt1998} suggest that the star formation is solely regulated by the amount of gas that exceeds a certain density threshold \citep{Gao2004, Lada2012, Evans2014, Zhang2014}. However, these simple scaling laws are constrained by the sample population size and break down over scales smaller than a few hundred pc, where the enclosed sample of molecular clouds decreases significantly \citep{Kruijssen2014}. 

Power-spectrum studies of giant-molecular-cloud maps in the Galactic disc have shown that the SFE and clump formation efficiency (dense gas mass fraction, DGMF) vary significantly on the scales of individual clouds, peaking at 10--30\,pc \citep{Eden2021}. This variation in SFE declines at a (smoothing) scale of 100\,pc. Furthermore, it was found that the distributions of SFE and DGMF in individual clouds are consistent with being lognormal \citep{Eden2012, Eden2013, Eden2015} and thus possibly a combination of several random factors, implying that extreme star-forming regions (or regions in which star formation is absent) are not necessarily due to special conditions. These results are also consistent with a simple Schmidt-Kennicutt law since the distribution of SFEs possesses a well-defined mean when averaged over kpc scales and a large number of clouds. Furthermore, the SFE/DGMF appears to vary by several orders of magnitude from cloud to cloud. Along with the nearly constant mean value of the distribution of SFEs, this fact suggests that differences between the individual clouds are more relevant to star formation than large-scale mechanisms such as density features, shear, and radial variations in metallicity. In particular, spiral arms appear to mainly only produce source crowding. \cite{Ragan2016} and \cite{Ragan2018} also confirmed no arm-associated signal in the fraction in the Hi-GAL catalogue of compact sources that are currently star-forming. 

These results agree with observations of spiral galaxies indicating that the \ce{H_2}/HI fraction and the SFE traced by infrared (IR) and ultraviolet (UV) emission in spiral arms are not significantly higher than in the inter-arm gas \citep{Kennicutt2003, GildePaz2007, Walter2008, Leroy2009, Obreschkow2009, Foyle2010}. Also, the fraction of GMCs formed from HI appear to be determined by the \ce{H_2} formation/destruction rate balance and stellar feedback \citep{Leroy2008}. These mechanisms act at small scales in the ISM. Except for starburst galaxies and ultraluminous IR galaxies (ULIRGs), internal radiative feedback is expected to determine the properties of molecular clouds with the minor influence of the external environment \citep{Krumholz2009}. These pieces of evidence challenge the idea that spiral arms may be direct triggers of star formation.

The problem of setting up a comprehensive model for SFE is further aggravated by the impact of large-scale radial changes in Galactic environments on the star-forming properties of the gas. The fraction of molecular gas has been observed to decrease rapidly with Galactocentric distance, from $\sim$100\,per\,cent within 1\,kpc to only a few per cent at radii greater than 10\,kpc \citep{Sofue2016}. Simultaneously, DGMFs peak at around 3--4\,kpc and then decline in the inner zone, where the disc becomes stable against gravitational collapse on large scales. This is the zone swept by the Galactic bar and star formation is suppressed for the life of the bar \citep{James2016}. The SFE, measured as either the integrated infrared luminosity from young stellar objects (YSOs), or the numbers of HII regions, per unit molecular gas mass, appears to be low but steady on kiloparsec scales at radii greater than 3\,kpc. Recent studies of the dense gas fraction within Galactic-plane molecular clouds suggest that the SFE increases with distance at radii greater than 5\,kpc \citep{Urquhart2021}.  
The SFE declines abruptly in the Central Molecular Zone (CMZ) within 0.5\,kpc \citep{Longmore2013, Urquhart2013}. This significant difference may be related to higher turbulent gas pressure in the CMZ, which raises the density threshold for star formation \citep{Kruijssen2014}, but the cause of such differences and transitions between these regions remains unexplained. 

The low SFE in the CMZ cloud G0.253+0.016 appears to be caused by a prevalence of shear-driven solenoidal (divergence-free) turbulence modes, in contrast to spiral-arm clouds, which typically have a significant compressive (curl-free) component \citep{Federrath2016}. A similar analysis of the Orion B molecular cloud \citep{Orkisz2017} finds that the turbulence is mostly solenoidal, consistent with its low SFR, but is position-dependent within the cloud, motions around the main star-forming regions being strongly compressive.
Thus, this significant inter-cloud variability of the compressive/solenoidal mode fractions may be a decisive agent of variations in the SFE. The SFE may also be affected by cloud collisions, which should produce highly compressive gas flows. 

The \ce{^{13}CO}/\ce{C^{18}O} ($J=3 \rightarrow 2$) Heterodyne Inner Milky Way 
Plane Survey (CHIMPS, \citealp{Rigby2016}) has produced a large sample of molecular clouds and the first large-scale map of molecular-gas temperatures. The high resolution of this survey reveals significant arm structures in more detail and greater contrast than similar surveys. Contrary to theoretical predictions \citep{Kruijssen2014}, the study of CHIMPS clouds revealed SFE is neither linked to turbulent pressure or Mach numbers in the disc \citep{Rigby_thesis}.

Together, these findings emphasise the need for the detailed analysis of large samples of molecular clouds from different regions in the Galaxy, relating their internal and external environmental conditions to their SFE and DGMF, as the next step in understanding the physics of star formation. 

This article presents the first full sample study of the turbulent modes and their relation to SFE in Galactic-plane clouds, thus testing the hypothesis that the SFE depends on the ratio of
solenoidal to compressive turbulence within clouds. This has already been suggested for
the G0.253+0.016 cloud \citep{Kaufmann2013, Johnston2014}  and is thought to be consistent with the assumption that the majority of power in SFE variations is concentrated on cloud scales. The sample considered consists of 2266 \ce{^{13}CO} sources extracted from the CHIMPS survey. The article is organised as follows. Sections \ref{survey} and \ref{data} briefly describe CHIMPS and the construction of the catalogue employed in the analysis. 
In Section \ref{methods}, we present the main concepts and implementation of the statistical method devised by \cite{Brunt2014} to estimate the solenoidal fraction $R$ from the observed data.
The results are described in Section \ref{results}, and discussed in Section \ref{conclusions} with particular emphasis on the relations between the solenoidal fraction and star-formation efficiency within clouds and distribution of clouds with the Galactocentric distance in the different Galactic environments covered by CHIMPS.

In appendix \ref{the_field_size} we examine the influence of the size of the field (see Section \ref{implementation}) on the calculation of the solenoidal fraction.

\section{Survey}
\label{survey} 

The \ce{^{13}CO}/\ce{C^{18}O} ($J=3-2$) Heterodyne Inner Milky Way Plane 
Survey (CHIMPS)  is a spectral survey of the  $J = 3 \rightarrow 2$ 
rotational transitions of \ce{^{13}CO} at 330.587 GHz and \ce{C^{18}O} at 
329.331 GHz. The survey covers $\sim$19 square degrees of the Galactic plane, spanning longitudes $l$ between  $27\fdg 5$ and $46\fdg 4$ and latitudes $|\, b \,| < 0\fdg5$, with angular resolution of 15\,arcsec. The observations were made over a period of 8 semesters (beginning in spring 
2010) at the $15$-m James Clerk Maxwell Telescope (JCMT) in Hawaii. Both isotopologues were observed concurrently \citep{Buckle2009} using the Heterodyne Array Receiver Programme (HARP) together with the Auto-Correlation Spectral Imaging System (ACSIS). The data obtained are organised in position-position-velocity (PPV) cubes with 
velocities binned in 0.5\,km\,s$^{-1}$ channels and a bandwidth of 
200\,km\,s$^{-1}$\,. The Galactic velocity gradient associated 
with the spiral arms (in the kinematic local standard of rest, LSRK) is matched by shifting the velocity range with increasing Galactic longitude. Varying the velocity range from $-50 < v < 150$\,km\,s$^{-1}$\,\ at $28^\circ$ to $-75 < v < 125$\,km\,s$^{-1}$\, at $46^\circ$, we recover the 
expected velocities of objects observed in the Scutum-Centaurus tangent, and the Sagittarius, Perseus and Norma arms. The \ce{^{13}CO} survey has mean rms 
sensitivities of $\sigma(T_A^{*})\approx 0.6$\,K per velocity channel, while for \ce{C^{18}O}, $\sigma(T_A^{*})\approx 0.7$\,K, where $T_A^{*}$ is the antenna temperature corrected  for atmospheric attenuation, ohmic losses inside the instrument, spillover, and rearward scattering \citep{Rigby2016}. 
The total column densities throughout the CHIMPS survey are estimated from the excitation temperature and the optical depth of the CO emission. The full calculation is outlined in \citet{Rigby2019}. Their method is a variation of the standard calculation of the excitation temperature and optical 
depth \citep{RomanDuval2010, Wilson2013} and uses \ce{^{13}CO}($J=3-2$) emission from CHIMPS and \ce{^{12}CO}($J=3-2$) emission from COHRS (\citealp{Dempsey2013}, Park et al.\, in preparation) at each position $(l,b,v)$ in the datacube on a voxel-by-voxel basis, under the assumption of local thermodynamic equilibrium.

\section{Data}\label{data} 

\subsection{Data preparation}\label{data_preparation}
Before proceeding with the cloud identification, the CHIMPS data are prepared following the recipe used by \cite{Rigby2019}. The reduced data are spatially smoothed to a resolution of 27.4 arcsec (resulting from the application of a 3-pixel FWHM Gaussian filter) to increase the signal-to-noise ratio (SNR). The smoothed data have rms values of $0.09^{+0.03}_{-0.03}$ K per 0.5 km\,s$^{-1}$ channel. This value is the median of the distribution with uncertainties corresponding to the first and third quartiles \citep{Rigby2019}. Because of the variable weather conditions and the varying number of active receptors during the four years of observations, the original CHIMPS datacubes do not present a completely uniform sensitivity across the entire survey \citep{Rigby2016}. To avoid loss of good signal-to-noise sources in regions of low background and to prevent high-noise regions from being incorrectly identified as clouds, the source extraction is performed on the SNR cubes instead of brightness-temperature cubes. An SNR map is created from an existing brightness temperature cube by dividing it by the square root of its variance component. The resulting data array measures the SNR at each voxel of the original cube. These operations are performed by specific packages in the JCMT Starlink suite \citep{starlink}. This approach was applied to continuum data
in the JCMT Plane Survey (JPS) by \citet{Moore2015} and \citet{Eden2017}, who noted that this method produced the best extraction results. Finally, the background noise is identified and subtracted from the SNR
cubes by applying the Findback filter with a set neighbourhood with a side of 50 voxels.

\subsection{Cloud extraction}\label{cloud_extraction}

To identify molecular clouds in the CHIMPS \ce{^{13}CO} data we employ the Spectral Clustering for Interstellar Molecular Emission Segmentation (SCIMES) algorithm \citep{scimes}.
In SCIMES, the global hierarchical structure within a molecular-line datacube is encoded into a dendrogram. 

The input parameters that define the emission dendrogram are taken as multiple of the background $\sigma_\textrm{rms}$. For signal-to-noise cubes, $\sigma_\textrm{rms}=1$ by definition. 
For each region, the SCIMES parameters are set to generate an emission dendrogram in which emission below $5\sigma_\textrm{rms}$ ($\mathtt{min\_val }=5\sigma_\textrm{rms}$) is not considered. This minimum SNR value for a feature to be detected as a source was chosen to mitigate the occurrence of false positives (artefacts arising at low noise levels). Each branch of the dendrogram is defined by an intensity change of $5 \sigma_{\textrm{rms}}$
($\mathtt{min\_delta}= 5 \sigma_{\textrm{rms}}$). This value is chosen to match $\mathtt{min\_val}$ so that two adjacent peaks are considered distinct only if the difference in their values is also greater than 5. In addition, the minimum number of voxels an emission peak must contain to be included in the dendrogram ($\mathtt{min\_npix}$) is set to 16, which is at least three resolution elements worth of voxels. This value corresponds to the volume of a cubic source with a width of 2.5 voxels in each of the three axes. Lowering this threshold increases the likelihood of identifying spurious noise artefacts as features of the emission. These specific values were chosen to match the corresponding {\sc FellWalker} configuration parameters used by \citet{Rigby2016} for their CHIMPS emission extraction.  Full details of the extraction will be published in a separate paper.

Since the distances to the dendrogram structures are not known, the volume and luminosity affinity matrices required for spectral clustering cannot be generated from spatial volumes and intrinsic luminosities. Instead, PPV volumes and integrated intensity values are used \citep{scimes, Colombo2018}. The complete emission extraction yields a catalogue of 2944 sources.

\begin{figure}
	\includegraphics[width=\columnwidth]{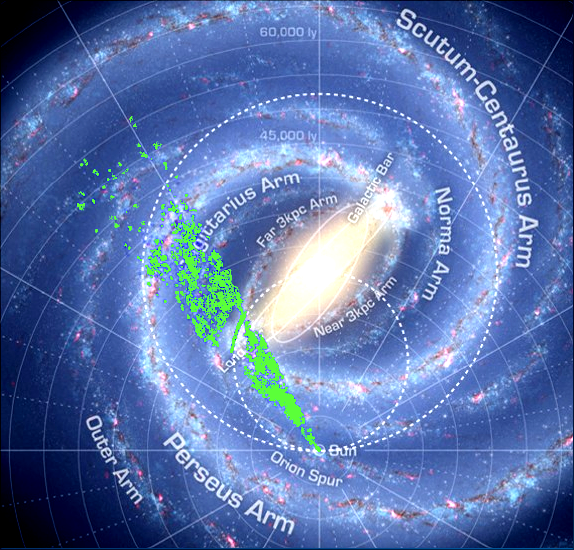}	
	\caption{Top-down view of the locations of the \ce{^{13}CO} (3 - 2) extracted through the
	SCIMES algorithm from CHIMPS. The background image is published by \citep{Churchwell2009}.The
	Solar circle and the locus of the tangent points have been marked as dashed lines.}
	\label{galmap} 
\end{figure}

To clean the catalogue of spurious sources and noise artefacts that are left after extraction, an additional filter is applied. This mask leaves those clouds that either extend for more than 8 pixels in each spatial direction or cover at least 3 velocity channels. This requirement ensures 
that each cloud is fully resolved in each direction (the width of the telescope beam being 2 voxels) and that the selection does not include sources with too small a field size for which 
the Fourier transform would not yield reliable information (see Appendix~\ref{the_field_size}).

In addition, clouds in contact with edges of the field of observation and those with no known column densities are removed from the catalogue. Masking the SCIMES catalogue with these requirements leaves 2266 sources. 
Distances are assigned using the ATLASGAL \citep{Urquhart2018} and CHIMPS \citep{Rigby2019} catalogues through a novel algorithm that performs
area searches to find the closest known sources to each SCIMES centroid. If this search returns multiple clouds, the distance that most sources have in common is chosen. If the distances in the set vary significantly, ATLASGAL clusters are checked, and cluster distances are assigned. For unassigned sources, a similar search is repeated considering CHIMPS sources that lie within SCIMES objects. Finally, Reid's Bayesian calculator \citep{Reid2014} is employed to estimate the distances of the remaining SCIMES sources with undetermined distances with a near-far probability of $0.5$). Fig.~\ref{galmap} shows the positions of the extracted sources superimposed on a sketch of the structure of the Milky Way.

The smallest clouds in this selection are large enough to include an envelope of rarefied gas around the densest, brightest peaks. This supports our
assumption of considering  \ce{^{13}CO} ($J=3\rightarrow 2$) to be optically thin in diffuse regions
\cite[with optical depth increasing around the peaks of
emission, where the cloud is densest, ][]{Rigby2016}. In a typical cloud, the volume occupied by the diffuse component far exceeds the denser parts. Through the CHIMPS optical depth maps prepared by \cite{Rigby2019}, we find that 97.7\,per\,cent of the voxels in the clouds identified by SCIMES with an associated optical depth have $\tau (^{13}{\rm CO}) < 1$. 

Although the selection criteria above allow us to consider a large number of sources, the spectral 
extent of some of them remains limited to 3 velocity channels. To ensure that these sources do not affect our analysis by introducing biases in the distribution of the physical quantities considered, we defined a subset of sources that span 8 velocity channels (thus allowing them to be larger than the volume contained in a cube with a side of 8 pixels). This sub-catalogue amounts to 954 entries. 

\begin{figure}
	\includegraphics[width=\columnwidth]{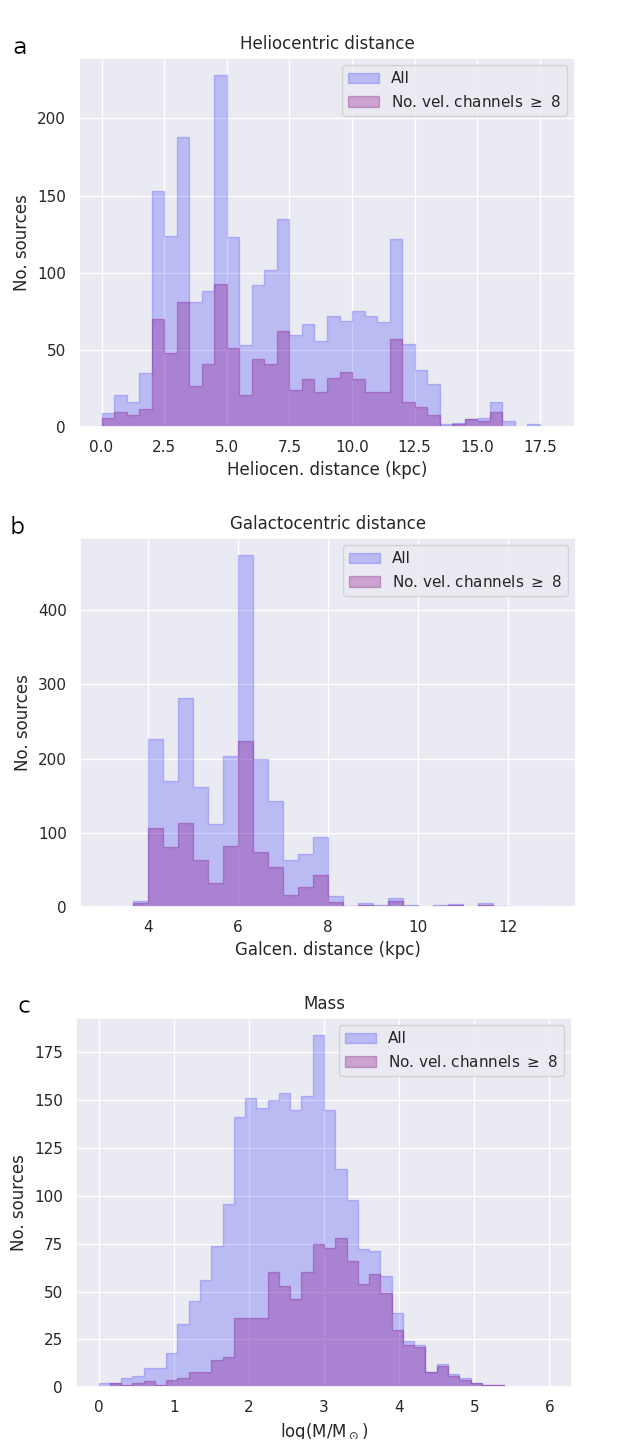}	
	\caption{Distributions of heliocentric distances (panel a), Galactocentric distances (panel b), and masses (panel c) of the \ce{^{13}CO} CHIMPS sources. The three panels show both the distributions of the entire sample extracted (2266 sources: blue) and of the subsample of clouds that span 8 or more velocity channels (954 sources: purple).}
	\label{distros} 
\end{figure}

With distances and velocity dispersions, masses, \ce{H_2} number densities and Mach numbers can be calculated. Fig.~\ref{distros} shows the distributions of distances and masses of both the full sample and the sub-sample of sources which span 8 or more velocity channels. The distribution of masses in the latter set has a higher mean.  This is likely to be a consequence of Larson's relations and approximate virialisation \citep{Larson1981}.

No sources closer than $3.5$\,kpc from the  
Galactic centre are found as the CHIMPS data do not probe sufficiently central longitudes. 
The sources in our sample reside within
the four main spiral arms, the Scutum-Centaurus, Sagittarius-Carina,
Perseus, and Outer arms and the smaller Aquila Rift and Aquila Spur features. Their distributions of Galactocentric distances reflect the arm structure. Fig.~\ref{distros}b displays large peaks at $\sim$4.5\,kpc and $\sim$6.5\,kpc. These are the locations of the Scutum and Sagittarius arms seen from the Galactic centre. The smaller peak at $\sim$7.5\,kpc corresponds to the Perseus arm. Part of the Scutum arm traverses the locus of tangential circular velocities and the sources in this area become clustered along this locus leaving gaps on either side (Fig.~\ref{galmap}).  
We note that this artefact originates from sources that have velocities greater than the terminal velocity due to non-circular streaming motions, which get binned at exactly the tangent distance, resulting in the apparent ‘gap’ and arc of sources lying on the tangent circle.

\subsection{Star formation efficiency}\label{star_formation_efficiency}

Star formation efficiency (SFE) can be understood as the rate of production of stars per unit mass of \ce{^{13}CO} (3-2)-traced clouds/clumps, integrated over some time scale. The star-formation history of a molecular cloud can be viewed as the luminosity of Young Stellar Objects (YSO) produced as a function of time. In this framework, SFE is quantified as the ratio of the IR luminosity of the YSOs embedded in a cloud to the mass of the cloud:

\begin{equation}\label{SFE_definition}
\mathrm{SFE} = \frac{L_\mathrm{star}}{M_\mathrm{cloud}}  
=\frac{1}{M_\mathrm{cloud}}\int_0^t\frac{dL}{dt} dt, 
\end{equation}

\noindent
where $dL/dt$ is the instantaneous star formation rate (SFR) in terms of the integrated luminosity $L$ of YSOs. Thus, both a long time scale and a high SFR can result in increased values of $L/M$. To uniquely identify $L/M$ with the SFE thus requires assuming $dL/dt$ to depend linearly on $dM/dt$. This assumption in turn entails that the stellar IMF be invariant and fully sampled in all star-forming regions, up to the maximum stellar masses \citep{Weidner2006}. An IMF that is filled stochastically \citep{Elmegreen2006}, may cause $L/M$ not to depend on the SFE linearly. In this case, an increase in the SFE still corresponds to an increase in $L/M$. The formation of larger star clusters with more fully sampled IMF and lager maximum stellar mass in larger clouds may also increase the observed $L/M$.  For clusters, $L$ is proportional to $M^2$ where $M$ is the cluster mass and so $\propto M_{\rm cloud}$ if the SFE is the same. 
This potential variation in the relationship between $L/M$ and SFE cannot be resolved by observations unless it is possible to distinguish every single star in the cluster. 

In theory, $L/M$ evolves with time (increasing $L$ and decreasing $M$) and it becomes
necessary to define the SFE in terms of a specific time-scale \cite[e.g., free-fall time, see][]{Cheavance2020}. However, since SFE is generally lower than 30\,per\,cent \citep{Lada2003}, we can assume that $M$ remains constant over the time-scales typical of star-formation observed in the mid and far IR continuum. Also, the stage of massive star formation 
that can be detected in the mid and far IR lasts for only hundreds of thousands of years \citep{Davies2011, Mottram2011b}, a short enough time to allow us to consider $L/M$ as a snapshot of the current or instantaneous SFE. 

The SFE of the CHIMPS sources can be estimated by assigning a luminosity to each source. 
We use luminosity and flux data from the {\em Herschel} InfraRed Galactic Plane Survey \cite[Hi-GAL][]{Elia2017}. Hi-GAL is a large-scale survey of the Galactic
plane, performed with the {\em Herschel Space Observatory} in five infrared continuum bands between 70 and 500\,$\mu$m.
Luminosity assignments are made using the integrated bolometric fluxes of the Hi-GAL sources contained within each SCIMES cloud. Since the Hi-GAL catalogue does not include velocity information, a Hi-GAL source is matched to a SCIMES cloud when its Galactic coordinates lie within the projection of the SCIMES cloud on the Galactic plane. This assignment is not always unique as projecting along the spectral direction may result in the full or partial overlapping of multiple SCIMES-extracted clouds. The position of a Hi-GAL source on the Galactic plane may thus belong to several distinct projected clouds. When this happens, the assignment is made unique by associating a Hi-GAL source with the SCIMES cloud that has the brightest \ce{^{13}CO}  (3-2) intensity along the spectral direction at the source's coordinates \citep{Urquhart2007}. This method allows us to define a luminosity for 1403 clouds in the original sample.

\section{Methods}
\label{methods}

\subsection{Principles}\label{principles}
 
Our turbulence analysis is based on the 
the statistical method developed by \cite{Brunt2010} and \cite{Brunt2014}, which allows
us to quantify the relative fraction of the solenoidal and compressive
turbulence modes present in a molecular cloud from emission and column density observations. The main idea behind the method is to reconstruct the
properties of a three-dimensional source from the
information contained in its observed two-dimensional line-of-sight projection. Assuming that the observed source is described by the three-dimensional field $\mathbf{F}$, its two-dimensional projection (average along one axis, the z-axis in this case) is denoted by $\mathbf{F}_p$.  It can be shown that the Fourier transform $\tilde{\mathbf{F}}_p$ of  $\mathbf{F}_p$ is
proportional to the $k_z = 0$ cut of the transform 
$\tilde{\mathbf{F}}$ of $\mathbf{F}$, 

\begin{equation}\label{fourier_cut}
    \tilde{\mathbf{F}}_p (k_x, k_y) \propto \tilde{\mathbf{F}}(k_x,
    k_y, k_z = 0).
\end{equation}

If $\tilde{\mathbf{F}}$ and $\tilde{\mathbf{F}}_p$ only
depend on the wavenumber $k = |\mathbf{k}|$ (isotropic
fields), the average properties of $\mathbf{F}$ can be
derived from their two-dimensional counterparts of
$\mathbf{F}_p$ through symmetry arguments. When a field such as the velocity or the momentum is measured in
observations, only its line-of-sight component is available.
A two-dimensional projected field is recovered by considering the Helmholtz decomposition of the line-of-sight component. 
According to the Helmholtz theorem, a vector field can be split into a  divergence-free (solenoidal or transverse) 
component, $\mathbf{F}_\perp$ and curl-free (compressive or
parallel) component, $\mathbf{F}_\parallel$. 
In Fourier space, the solenoidal and compressive components are linked through (local) orthogonality. As the name
suggests, the divergence-free (solenoidal) component encodes
the turbulent, vorticose modes of a flow. Compressive modes,
accounting for compression and expansion of the gas are
embodied by the curl-free component. These modes are likely
to be connected to star-formation. 

To obtain a unique
decomposition, the vector field must satisfy suitable
boundary conditions (the Helmholtz decomposition is defined
up to a vector constant). In
particular, it is required that the field should decay to zero smoothly on
the boundary. This condition also ensures that the Fourier
transforms of the observed field are well-behaved as these
fields are not naturally periodic. Isolated, gravitationally
bound molecular clouds possess a natural boundary, however,
when the signal is truncated artificially by the edges of the
observed field, apodisation of the emission at the edge is
required to restore a suitable boundary. 

As mentioned above, statistical isotropy is also required for
the method to be applied. Sources of strong anisotropy such
as strong magnetic fields or filamentary shapes thus heavily
affect the reliability of the results. Fields with steep power spectra should also be avoided. In practice, such power spectra show high sensitivity to low
spatial frequencies which are poorly sampled statistically (see also Section \ref{power_spectra}).
Assuming the emission line under consideration is optically
thin and that the emissivity depends solely on the volume 
density, the PPV datacube can be translated into a density-weighted field spanning the region of observation. This field is the `momentum density',

\begin{equation}\label{m_density}
\mathbf{p} = \rho \mathbf{v}, 
\end{equation}

\noindent
composed of the volume density $\rho$ and the velocity field
$\mathbf{v}$. 

The ratio of the variance of transverse momentum density to
the variance of the total momentum density gives the
solenoidal fraction, $R$. This fraction represents the amount
of power in the solenoidal modes of the momentum density in a given region of space and can  be  expressed as the ratio between the variances of the transverse (solenoidal) momentum and the variance of the total momentum, 

\begin{equation}\label{b1}
    R = \frac{\sigma^2_{p_\perp}}{\sigma^2_{p}}.
\end{equation}

\cite{Brunt2014} demonstrated that the solenoidal fraction
can be expressed in terms of observable quantities: the zeroth, first, second velocity moments, and their power
spectra. The first three velocity moments are defined as
 
\begin{equation}\label{b2}
    W_0 = \int I(v)\ dv, 
\quad
    W_1 = \int vI(v)\ dv,
\quad
    W_2 = \int v^2 I(v)\ dv
\end{equation}

\noindent
or their counterparts in a frame of reference set at the centre of mass of the molecular cloud. With the assumption that the thermal linewidth is negligible
compared to the overall velocity dispersion, the velocity moments can be recast in terms of density \citep{Brunt2014}:

\begin{equation}\label{b2_1}
    W_0 \propto  \int \rho(z)\ dz, 
\quad
    W_1 \propto  \int v(z) \rho(z)\ dz, 
\quad
    W_2 \propto  \int v(z)^2 \rho(z)\ dz . 
\end{equation}

\noindent
These moments allow for the solenoidal fraction to be written as
 
\begin{equation}\label{b3}
R = \Bigg[\frac{\langle W_1^2\rangle}{ \langle
W_0^2\rangle}\Bigg]\Bigg[\frac{\langle W_0^2 / \langle W_0
\rangle^2 \rangle}{1 + A(\langle W_0^2\rangle/ \langle W_0 \rangle^2 -1)} \Bigg]
\Bigg[ g_{21} \frac{\langle W_2 \rangle}{\langle W_0 \rangle}
\Bigg]^{-1}B,
\end{equation}

\noindent
where 

\begin{equation}\label{b4}
     A = \frac{(\sum_{k_x}\sum_{k_y}\sum_{k_z}f(k))-f(0)}{\sum_{k_x}\sum_{k_y}f(k))-f(0)},
\end{equation}

\noindent
and

\begin{equation}\label{b5}
     B = \frac{\sum_{k_x}\sum_{k_y}\sum_{k_z}f_\perp(k)\frac{k_x^2 + k_y^2}{k^2}}{\sum_{k_x}\sum_{k_y}f _\perp (k)},  
\end{equation}

\noindent
with $f(k)$ and $f_\perp(k)$ being the angular (azimuthal) averages of the power spectra of the zeroth and first moments \cite[notation after ][]{Orkisz2017}.
The constant  $g_{21}$ is a statistical correction factor that accounts for
the correl
ations between the variations of $\rho$ and
$\mathbf{v}$ (if  $\rho$ and $\mathbf{v}$ are not correlated,
$g_{21} = 1$). In terms of density, velocity and the spatial average of
the density $\rho_0$, $g_{21}$ is expressed by the variance of
the three-dimensional volume density $\langle (\rho /
\rho_0)^2 \rangle$ as

\begin{equation}\label{b6}
    g_{21} = \frac{\langle\rho^2v^2\rangle/\langle 
    \rho^2\rangle}{\langle \rho v^2\rangle/\langle \rho
    \rangle} =  \Bigg \langle \frac{\rho^2}{\rho_0^2} \Bigg
    \rangle^\epsilon. 
\end{equation}

The exponent $\epsilon$ is a 
is a small positive constant which is the exponent of the power law expressing the relation between the variance of the velocity $\sigma_v^2$ and the density $\rho$ . The variance of the three-dimensional volume density  $\langle (\rho/\rho_0)^2 \rangle$ is estimated from the column density cubes following the method introduced by \cite[]{Brunt2010} with $\rho_0$ being the spatially averaged volume density.

In the hypersonic regime (Mach number $\mathcal{M} > 5$), the solenoidal fraction
becomes independent of the type of forcing and converges to
$R \sim 2/3$ \citep{Brunt2014}. This specific value reflects
the equipartition of momentum between the compressive and
solenoidal mode \citep{Fedderath2008}. Values of the
solenoidal fraction that are higher than $2/3$ imply that
the relative fraction of momentum density in solenoidal modes in
the flow exceeds that in compressive modes. Thus, star
formation tends to be suppressed.  A solenoidal fraction smaller than $2/3$ implies a loss of equilibrium in favour of the compressive modes of the flow. When this situation occurs, a cloud is more likely to form stars.

\begin{figure}
	\includegraphics[width=\columnwidth]{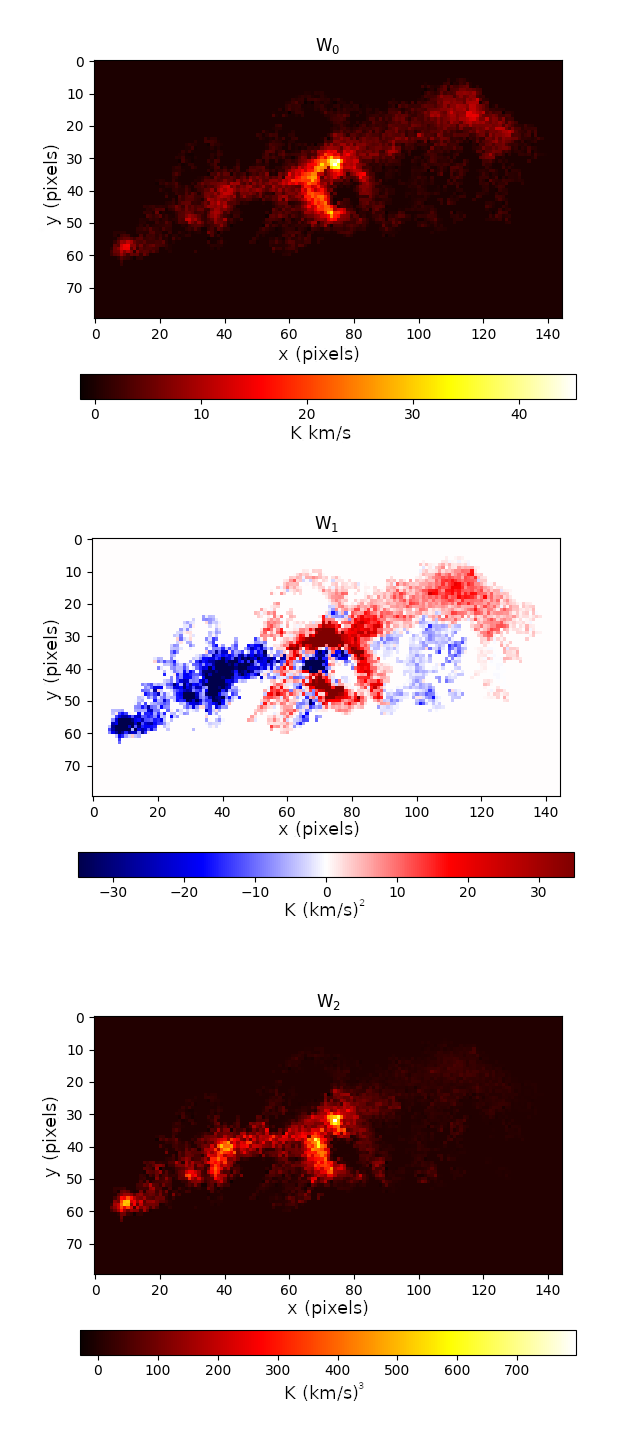}	
	\caption{Example of integrated intensity ($W_0$), first ($W_1$) and second moment ($W_2$) maps for a \ce{^{13}CO} emission source in CHIMPS.}
	\label{moments} 
\end{figure}

\subsection{Implementation}\label{implementation}

\subsubsection{Moments}\label{emission_moments}

The method described above is applied to our selection of SCIMES clouds extracted from the CHIMPS \ce{^{13}CO} (3-2) emission data.  The emission of each cloud in the selection is isolated via a mask constructed from the SCIMES clusters assignment. 
The velocity moments whose power spectra in Equation \ref{b3} must be calculated in the frame of reference of the centre of mass of the cloud. Thus, to express the velocity moments in the centre-of-mass frame, first the centroid velocity of the cloud in the LSR frame is calculated. This quantity is simply given by the ratio

\begin{equation}\label{com_velocity}
    V_c = \frac{\langle W_1^\mathrm{obs}\rangle}{\langle W_0
    \rangle}, 
\end{equation}

\noindent
of the spatial means of the first moment in the
observer's frame and $\langle W_1^\mathrm{obs}\rangle$
and of the zeroth moment $\langle W_0 \rangle$. Notice that, not being velocity-weighted, $W_0$ is invariant of the frame of reference. The resulting change of coordinates gives

\begin{equation}
    v = v_\mathrm{obs} - V_c
\end{equation}

\noindent
(adopting the same notation as before).  Finally,
substituting in the first and second moments yields

\begin{equation}
    W_1 = \int (v_\mathrm{obs} - V_c) I(v_\mathrm{obs})
    \ dv_\mathrm{obs}
\end{equation}
 
 \noindent
and
 
\begin{equation}
    W_2 = \int (v_\mathrm{obs} - V_c)^2 I(v_\mathrm{obs})\
    dv_\mathrm{obs}.
\end{equation}

Once the moment maps of a cloud have been constructed, the cloud is extracted by
enclosing it into a square region of the map. Although the emission values
decline naturally to zero at the boundary of the cloud, to ensure that the
boundary conditions required by the method are respected, we introduce a
field size that depends on the size of the projected cloud. The field size
(side) of this region is determined by considering the maximum extension of the
cloud along the coordinate axes with an added 5-pixel padding in every
direction. 
We found that, in the case of the clouds in our sample, the correction in the
value of the three-dimensional variance of the momentum density
\citep{Brunt2010} to account for the added padding is negligible. It only
affects $R$ by a factor of $10^{-4}$.

\begin{figure}
	\includegraphics[width=\columnwidth]{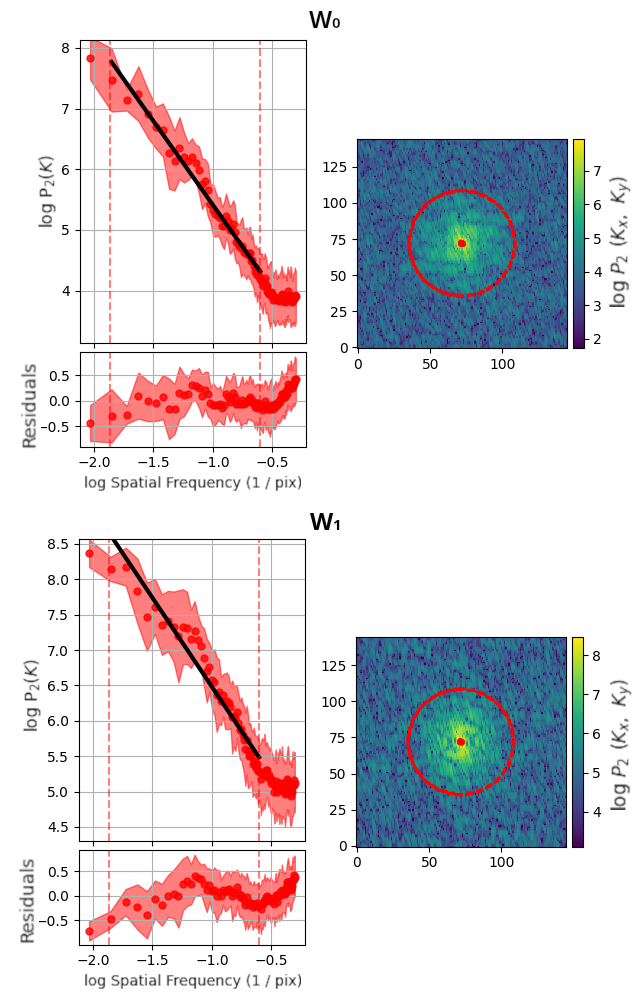}	
	\caption{Example of power spectra of the zeroth ($W_0$) and first ($W_1$) moment maps. Each panel shows both the angular averaged 1D and full 2D power spectra. The dashed lines in the one-dimensional spectra and the corresponding red circles in the two-dimensional power spectra delimit the region over which the spectrum is fitted with a segmented linear model. The fitted power-law model of the 1D spectrum is denoted by the solid black line.}
	\label{ps} 
\end{figure}

\subsection{Power spectra}\label{power_spectra}

The power spectra of the moments maps are calculated using the {\sc PowerSpectrum}
method in {\sc Turbustat} \citep{Koch2019a}, a Python package that
implements a suite of tools devoted to the statistical analysis of turbulence
\citep{Koch2019}. {\sc PowerSpectrum} implements a model for the computation of the
full two-dimensional spatial power spectrum of an image (an elliptical power-law
model). A radial profile of the two-dimensional power spectrum produces the
azimuthally averaged one-dimensional power spectrum that is required for the
calculation of the solenoidal fraction. {\sc PowerSpectrum} both provides an automatic correction for the telescope beam (deconvolution) and a power-law fit for the one-dimensional power spectrum (see Figure \ref{ps}). 

\begin{figure}
	\includegraphics[width=\columnwidth]{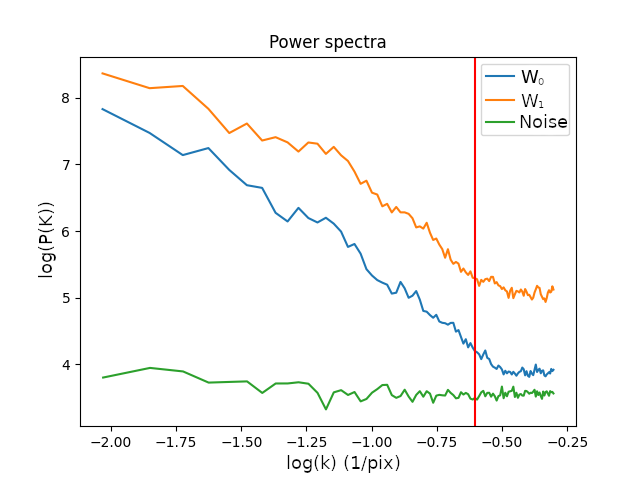}	
	\caption{A representative one-dimensional power spectrum of the noise and of the zeroth and first moment maps for CHIMPS clouds. The vertical red line denotes the frequency cut that corresponds to twice the resolution of the telescope. As the effects of noise and the correction for the beam may become more significant above this threshold, higher frequencies are discarded for the fitting of power spectra used in the calculation of the solenoidal fraction.}
	\label{psn} 
\end{figure}

To avoid large deviations of the power spectrum on small scales (high spatial frequencies) where the information has been lost by the spatial smoothing applied to the image (convolution of the beam),  only spatial frequencies that correspond to twice the FWHM value of the telescope beam are considered. This correction also accounts for the increase in power at high frequencies generated by the over-sampling of the beam \citep{Koch2019a}. 
A 2-beam frequency cut, corresponding to 4 pixels in CHIMPS, also mitigates the impact of the noise which is more severe at higher spatial frequencies (see Fig.~\ref{psn}).

Correlated voxel noise and systematic errors makes the noise in the emission maps differ from  Gaussian white noise.  To compute the power spectrum of the noise, we consider a region of the survey where the signal is absent and cut out a cube that corresponds to the size of each isolated cloud. This noise template is meant to reproduce a systematic behaviour over a finite number of channels. The power spectrum of the noise is then calculated as the average of the power spectra of the 2D map corresponding to each velocity channel, multiplied by the number of channels (to mimic the square root of the sum of squares that would correspond to the propagation of uncertainty when calculating the moment of the signal). Considering that both the zeroth and first moment of the signal are a linear combination of the channel maps, the same noise spectra can be compared to signal power spectra \citep{Orkisz2017}. 

Modelling the power spectra of the observable moments as the sum of the beam-convoluted signal spectrum and a noise spectrum \citep{Brunt2010, Orkisz2017}, we find that the amplitude of the noise component is several orders of magnitude smaller than the signal spectrum, becoming comparable in magnitude at frequencies around the telescope resolution (see Fig.~\ref{psn}).
However, the variation in the signal dominates noise at all scales. This is a consequence of extracting clouds in signal-to-noise cubes with a minimum SNR set to 5. In $W_1$ maps, the signal amplitude is amplified by 
multiplication with the velocity values. The noise component can thus be neglected in the fitting of the power spectra over frequencies below the 2-beam-width threshold. 
Although power laws alone may provide sufficiently good fits for the one-dimensional power spectra of some clouds, we encountered cases for which they are not enough to obtain an accurate fit over the entire spectrum. Thus, we chose to use Fourier-transformed data with a linear interpolation between the points to represent the power spectra. 
Interpolation supplies values of the power-spectrum model at all frequencies required in the summations in Equation~\ref{b3}. Data interpolation also mitigates larger uncertainties at low spatial frequencies caused by poor sampling at these frequencies.

\subsection{Density-velocity correlations}\label{density_velocity_correlations}

The exponent $\epsilon$ in Equation~\ref{b6} is set to $0.15$.  This value was derived by \cite{Orkisz2017} in their analysis of the solenoidal fraction in Orion B. Their estimation of
the relation linking local density and velocity
dispersion is based on several emission lines with different
spatial distributions in the mean spectrum (mean
line profiles). They considered five
isotopologues to trace gas at different densities: \ce{^{12}CO}($J = 1\rightarrow 0$) and \ce{HCO^{+}}($J = 1\rightarrow 0$) for low density gas \citep{Pety2017}, 
\ce{^{13}CO}($J = 1\rightarrow 0$) for the bulk of the cloud \citep{Orkisz2017},
 \ce{C^{18}O}($J = 1 \rightarrow 0$) for denser and shielded regions \citep{HilyBlant2005}, 
  and \ce{N_{2}H^{+}}($J = 1 \rightarrow 0$) for the densest cores \citep{Kirk2016}.
 \cite{Orkisz2017} devised an empirical relation between the fitted velocity dispersion velocities ($\delta v$) and lowest emission density ($\rho(\mathrm{H}_2)$) from the data of the five species:

\begin{equation}
    \delta v \propto \rho(\mathrm{H}_2)^{-0.15}.
\end{equation}

The slope 
$-\epsilon = -0.15$ is derived from a least-squares fit of the variation of the FWHM with the density. 
\cite{Orkisz2017} estimated that possible systematic errors in the \ce{^{12}CO} (1-0), \ce{HCO^+} (1-0), and \ce{N_2H^+} (1-0) densities and the \ce{^12CO} (1-0) and \ce{HCO^+} (1-0) linewidths tend to steepen the slope of the power law. Thus, $\epsilon = 0.15$ should be considered as an upper bound. We adopt this value for the correction factor $g_{21}$
(equation \ref{b6}). A lower bound of $g_{21}$ is provided by $\epsilon = 0.05$ as estimated by \cite{Brunt2014}.

\section{Results}
\label{results}

\subsection{The solenoidal fraction}\label{the_solenoidal_fraction}

The sample selection described in Section \ref{cloud_extraction}
produces a collection of 2266 SCIMES clouds for which the solenoidal fraction is calculated. While the selection requirements (Section \ref{data_preparation}) ensure that both sheet and filamentary structures are considered, a subsample selected by setting the spectral extension to a minimum of 8 voxels implies that these clouds are fully resolved in each direction (the width of the beam being 2 voxels) and comprise sizes that mitigate the impact of very steep density gradients on the calculation of the solenoidal fraction. This subsample amounts to 954 sources. In the text, we will refer to this set of sources as the velocity-limited sample. 

\begin{figure}
	\includegraphics[width=\columnwidth]{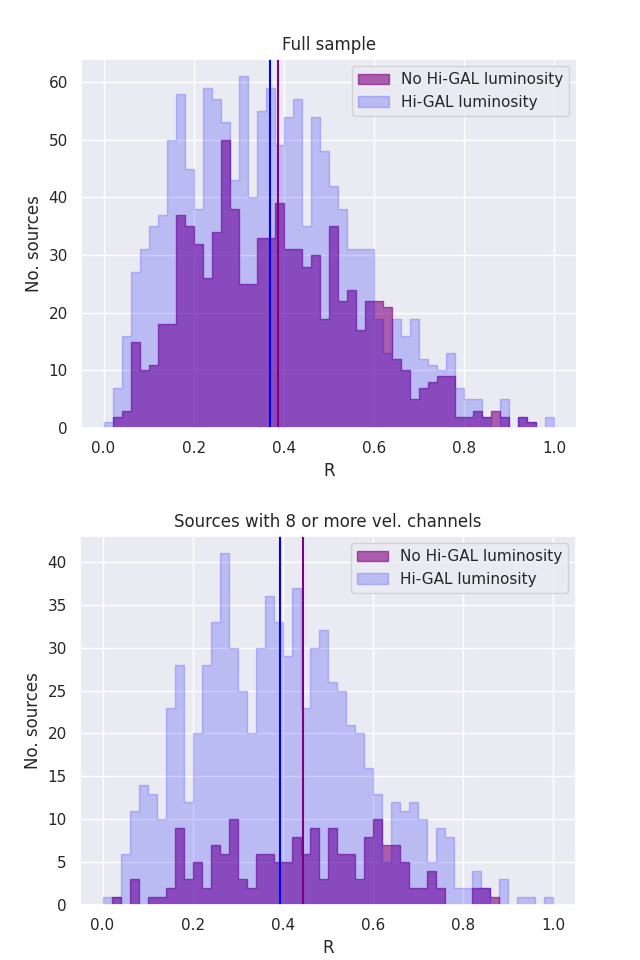}	
	\caption{Distributions of the solenoidal fraction within the full (top panel) and size constrained (bottom panel) samples of CHIMPS clouds. In both panels, the purple histogram traces the distribution of the subset of sources that do not have Hi-GAL luminosity counterparts. The vertical lines denote the means of the distributions.}
	\label{sol_lumo} 
\end{figure}

The solenoidal fraction $R$ (introduced in Section~\ref{principles}) is calculated through an algorithm that automates the steps described in subsections \ref{emission_moments}, \ref{power_spectra}, \ref{density_velocity_correlations}, allowing for the method to be applied to a large sample. 
This algorithm produces the value of $R$ associated with each cloud in a SCIMES cluster assignment map, given its corresponding cloud catalogue number produced by SCIMES, the survey emission map and the column-density data as input.

\begin{figure}
	\includegraphics[width=\columnwidth]{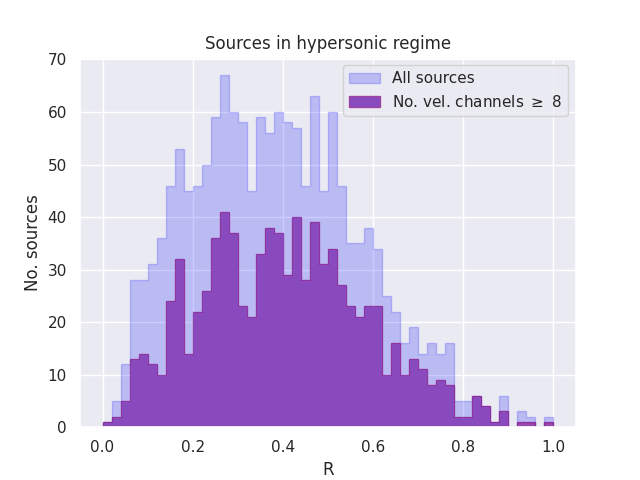}	
	\caption{Distributions of solenoidal fraction for clouds in hyper-sonic regimes (Mach number $> 5$).  
	This subsample comprises 67.9\,per\,cent (1538 sources) of the original selection for which the solenoidal fraction is calculated (2266). For the 
	set of sources that cover 8 or more velocity channels, the hypersonic clouds (865) amount to 90.7\,per\,cent of the sample (954). With solenoidal fractions $< 2/3$, the majority of hypersonic clouds have the potential to form stars. 
	}
	\label{hypersonic} 
\end{figure}

Fig.~\ref{sol_lumo} shows the distributions of $R$ for sources with and without associated Hi-GAL bolometric luminosities (see Section~\ref{star_formation_efficiency}) in both the full sample and the velocity-limited subsample. These distributions appear to show that the sample without associated luminosities is shifted to slightly higher solenoidal fractions. This behaviour, most evident in the velocity-limited subsample, is consistent with the hypothesis that a higher value of $R$ reduces the likelihood of star formation. To check for significance, a Kolmogorov-Smirnov test is performed over the two distributions of Fig.~\ref{sol_lumo}. Following the convention with the null hypothesis stating that the two sample distributions are drawn from the same population. 
The test returns   $k = 0.059$ with p-value $=0.0475$ for the full sample and  $k = 0.157$ with p-value $\ll 0.001$ for the velocity-limited sources, allowing us to reject the null hypothesis. 

Error estimation in the solenoidal fraction was performed by comparison between the original catalogue and a further calculation on emission maps perturbed by the addition of the square root of the corresponding variance maps. The method returned an average error of 8.6\,per\,cent. When taking into account the uncertainties on $\epsilon$ the error rises to $\sim$15\,per\,cent. Both values are consistent with the 8--13\,per\,cent range found for the Orion B emission \citep{Orkisz2017}. \

Isolating the subset of sources in hypersonic regimes\footnote{Mach numbers are calculated as the ratio of the non-thermal and thermal components of the dispersion velocity as defined in \cite{Rigby2019}} reveals (see Fig.~\ref{hypersonic}) that this selection comprises 67.9\,per\,cent of the full sample and 90.7\,per\,cent of the velocity-limited sample. 

In turn, only 5.4\,per\,cent and 8.1\,per\,cent of the hypersonic sources in the full and velocity-limited sets, respectively, have $R > 2/3$. Most of the selected clouds thus have the potential to form stars. Values of $R$ that exceed $2/3$ may be caused by systematics and measurement errors \cite[the estimation of Mach numbers, for instance, relies on excitation temperatures from the CHIMPS catalogue that may have issues related to non-LTE conditions, ][]{Rigby2019}.
These fractions also imply that the result is free of potential concerns over the nature of the forcing mechanism being a factor in the value of the solenoidal fraction. 

At these sonic regimes (hypersonic Mach number), complete mixing of turbulent modes is expected \citep{Federrath2011, Brunt2014}, so that the momentum equipartition would yield $R = 2/3$ with a fraction of  compressive modes equal to $1-R = 1/3$. Variations from these ratios can either indicate a specific forcing for the turbulence or the presence of an ordered flow which is superimposed on top of the turbulent flow \citep{Brunt2014}. 
As a specific forcing for the turbulent flow is more likely at transonic Mach numbers ($0.8 < \mathcal{M} <1.2$), the small fraction of CHIMPS clouds at transonic velocities implies that the forcing mechanism does not appear to be a factor in determining the solenoidal fraction for this sample. The solenoidal fraction is thus more likely to be set by the superimposed ordered flow (collapse or outflow resulting from star formation, the combination of unresolved compressive motions, or other kinds of velocity gradients along the line of sight).

\begin{figure}
	\includegraphics[width=\columnwidth]{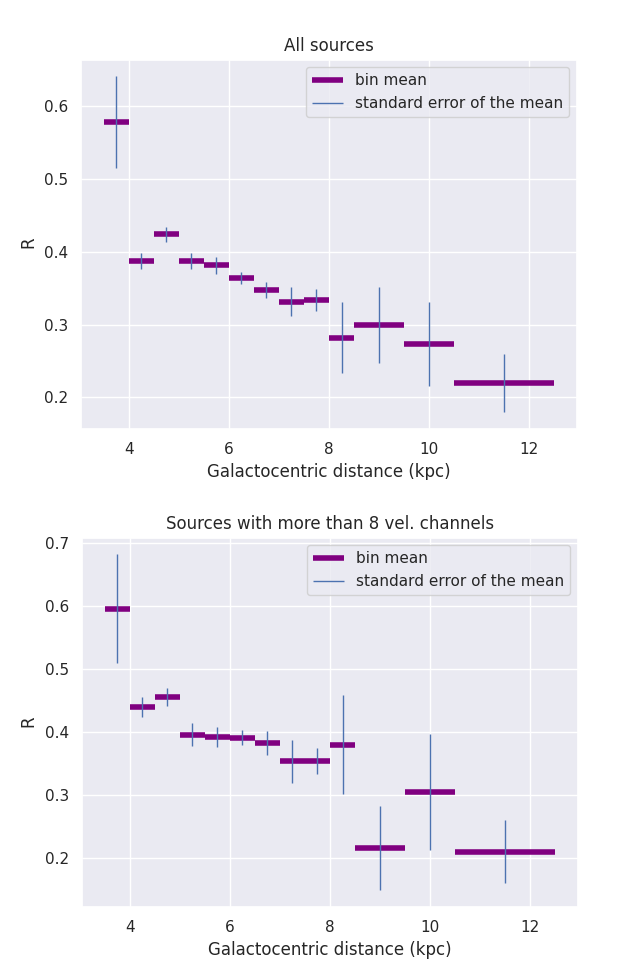}	
	\caption{Distributions of the solenoidal fraction
	with Galactocentric distance. The size of the bins is
	adjusted to the number of sources, being 0.5-kpc wide until 8\,kpc and 1-kpc wide from 8.5 to 10.5\,kpc. At distances larger than 10\,kpc, clouds are collected in a single 2-kpc bin. The horizontal purple lines
	indicate the mean value within each the bins. The vertical bars represent
	the standard error of the mean.}
	\label{poster} 
\end{figure}

Fig.~\ref{poster} shows the distribution of mean solenoidal fraction with Galactocentric distance. The width of the bins is 0.5\,kpc up to 8.5\,kpc radius, 1\,kpc from 8.5 to 10.5\,kpc and 2\,kpc past this distance. The reason for using irregular bin widths is to reduce biases by considering bin populations of similar sizes. Bin widths are represented by the length of the horizontal lines that indicate the mean value of the solenoidal fraction in each bin.  The solenoidal fraction peaks at the 3--4-kpc bin. This result requires confirmation by the analysis of a further sample at lower longitudes, but may be consistent with the disc becoming stable against gravitational collapse at these radii. Observations of such a sample are underway as part of the follow-up survey CHIMPS2 \citep{Eden2020}. 

The number of clouds with distances smaller than 4\,kpc amounts to 9 sources in the full sample and 6 in the velocity-limited set. The clouds in both sets have project sizes (number of pixels of their projection) ranging from 117 to 3202 pixels (with an average of 940) and field sizes from 21 to 108 pixels (the two sets have common extremals). Both sets include two clouds with field sizes above 85 pixels, see Appendix \ref{the_field_size}). These clouds do not present any special, unique features related to the size of their fields and are consistent with the entire population. The calculation of the their solenoidal fractions is thus free from biases linked to specific field sizes. Visual inspection of their size distribution is supported by the  Kolmogorov-Smirnov test result ($k= 0.46$ and p-value $=0.03$), suggesting that these clouds are sampled from the full distribution. The small size of the set makes this a point of low significance but nonetheless invites further work at low Galactic longitudes.

The solenoidal fraction then declines with a shallow gradient with
increasing Galactocentric distance. For Galactocentric distances
greater than 4\,kpc, a Spearman test returns $r=-0.15$ (for the full
sample, and $r=-0.19$ for the velocity-limited sources) with a p-value
$\ll0.001$, indicating that the solenoidal fraction declines with distance from the Galactic centre.
This decrease corresponds to a shallow gradient with a slope of $-0.02$\,kpc$^{-1}$
with no signal present at the spiral-arm radii (4.5, 6.5 and 7.5\,kpc
seen in Fig.~\ref{distros}). This result is in agreement with previous studies that found no significant arm associated signal in star-formation-related observational parameters \citep{Ragan2016, Ragan2018}. To check if this result arises from distance biases, i.e. we are more sensitive to the solenoidally-dominated envelopes of clouds when they are located closer to the observer, we construct a distance-limited subsample. This set only includes sources with Heliocentric distance between 8 and 12\,kpc and contains 581 sources with Galactocentric distances ranging from 3.9 to 8.1\,kpc. A Spearman test recovers the negative $R$-Galactocentric distance correlation with $r = -0.21$ with p-value $\ll 0.001$.

No significant correlation (Spearman statistics) was found between the solenoidal fraction, mass, and Mach number. In particular, the solenoidal fraction is not correlated to the volume of the clouds (number of voxels) ensuring that the results are not affected by resolution biases. 

These results suggest that the state of the physical properties of a cloud and thus its likelihood to form collapsing cores may be linked to the Galactic environment or individual cloud formation histories
in which the cloud is located, slowly changing in the disc and possibly steepening into the bar-swept region and continuing into the CMZ which has very low SFE \citep{Longmore2013, Urquhart2013}. In this picture the spiral arms are not a strong influencing factor.

\subsection{Star formation efficiency}\label{SFE2}

\begin{figure*}
	\includegraphics[width=\textwidth]{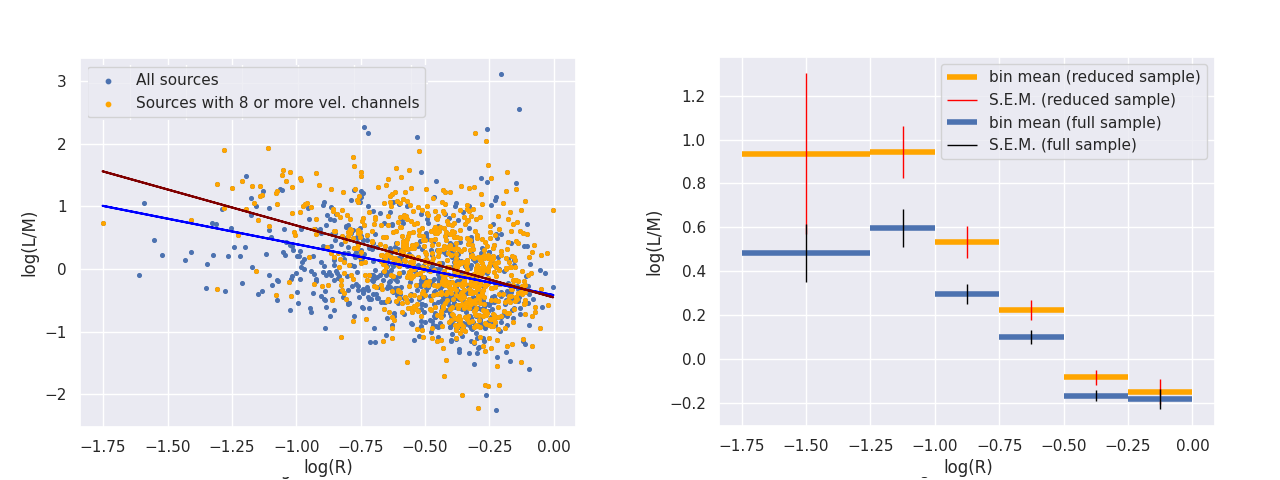}	
	\caption{Star formation efficiency defined as $L/M$ (in units of Solar mass and Solar luminosity) as a function of the solenoidal fraction. The  IR continuum luminosity from the YSOs and the masses of the sources (CO mass) are derived from independent measurements and can be considered largely independent variables. Left panel: Orange dots represent the sub-sample with 8 or more velocity channels.  The blue and brown solid lines are weighted linear fits to the full and reduced samples, respectively. The weights are the standard deviations of the $L/M$ distribution within solenoidal fraction bins with width 0.1. Right panel: Distributions of the SFE with the $R$ bins for the velocity-limited (reduced) sample (orange) and full sample (blue). The horizontal lines indicate the mean value within each of the bins. The vertical bars represent the standard error of the mean (S.E.M.). }
	\label{sf} 
\end{figure*}

We now consider the relation between the star formation efficiency as defined in Section \ref{star_formation_efficiency} and the solenoidal fraction. Fig.~\ref{sf} shows a negative correlation (Spearman $r = -0.33$, p-value $\ll 0.001$ for the full sample and $r = -0.37$, p-value $\ll0.001$ for the velocity-limited sources) is found between SFE and $R$. This correlation is again consistent with the hypothesis that star formation is more likely to occur in clouds with more power in the more dominant compressive turbulent modes. 

\begin{figure}
	\includegraphics[width=\columnwidth]{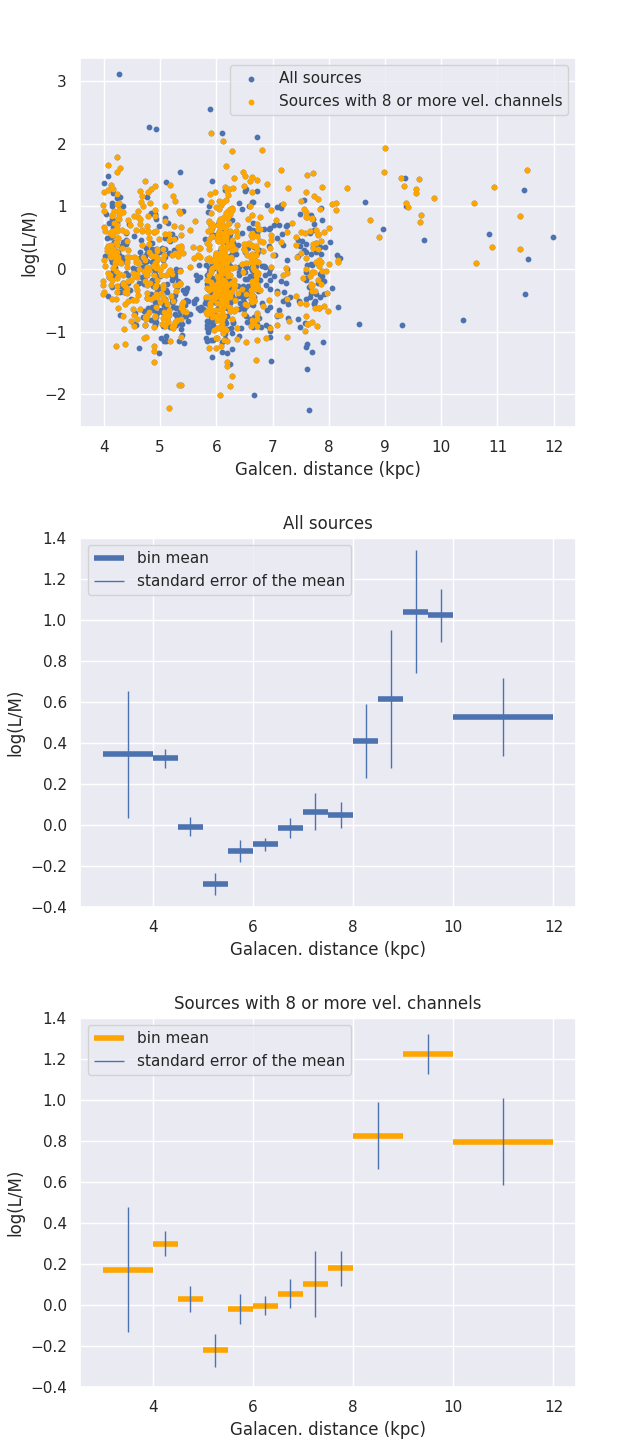}	
	\caption{Distributions of the SFE with Galactocentric distance. A scatter plot of the full (blue) and reduced (orange) sample is shown in the top panel. In the mid and bottom panels, the size of the bins is
	adjusted to the number of sources. The bins are 0.5\,kpc wide until 8\,kpc and 1-kpc wide from 8 to 10\,kpc for the velocity-limited sample (until 10\,kpc for the full). At distances larger than 10\,kpc, clouds are collected in a single 2-kpc bin. The horizontal lines
	indicate the mean value within each of the bins. The vertical bars represent the standard error of the mean.
}
	\label{semilog} 
\end{figure}

To ensure that the $L/M-R$ relation is not affected by distance or completeness biases, we consider a distance-limited subsample (581 sources with Heliocentic distances between 8 and 12\,kpc) and test the $L/M-R$ correlation within this set. A Spearman test returns $r = -0.36$ with p-value $\ll 0.001$,  indicating that the trend is still present.

We show the distribution of SFE as a function of the distance from the Galactic centre in Fig.~\ref{semilog}.
The distribution appears to show a positive correlation with increasing Galactocentric distance from 5 kpc (Spearman $r \simeq 0.22$ with p-value $\ll 0.001$ for both the full sample and the velocity-limited set). Although this measurement shows a significant scatter and may be biased towards clouds associated with more luminous star formation with increasing distance, this result is in agreement with the shallow gradient by which the average solenoidal fraction decreases with distance from the Galactic centre, except for the 3-5\,kpc range, where both solenoidal fraction and SFE increase inwards. As we have seen above, this region below 4\,kpc is poorly sampled by CHIMPS and the clouds found here do not present any distinctive geometric features. Although the reasons for this are unclear, the higher SFE observed between 4 and 5\,kpc may be attributed to higher emission from star-forming cores enveloped by large regions of rarefied gas that contributes to their high solenoidal fractions. 
Again there is no evidence for significant localised increases in the SFE corresponding to spiral arms. These findings agree with the SFE calculated for the Galactic Ring Survey \cite[GRS, ][]{Dib2012} and the SEDIGISM-ATLASGAL dataset \citep{Urquhart2021}. 

\begin{figure}
	\includegraphics[width=\columnwidth]{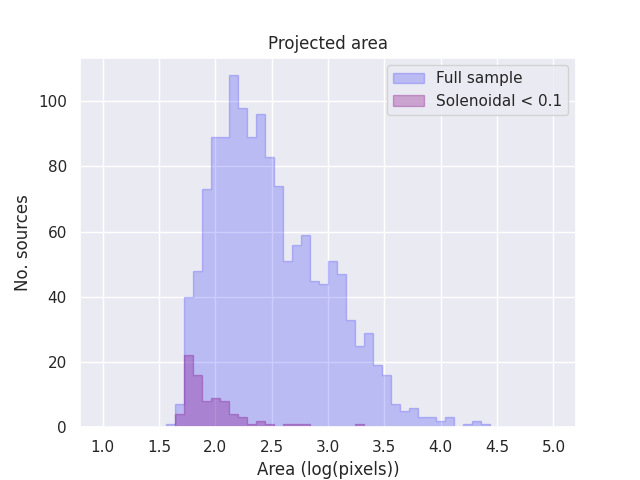} 
	\caption{Distributions of the projected areas of clouds with solenoidal fraction < 0.1 (purple) compared to the full sample (blue).}
	\label{upper_left} 
\end{figure}

To check for potential biases in the SFE-solenoidal-fraction relation that may originate from the application of Fourier transforms on fields of small size (which is not likely to yield useful information on the turbulent modes in the cloud), we first check the correlation between field size and solenoidal fraction. A Spearman test reveals a positive correlation (full sample: $r = 0.53$, p-value $\ll 0.001$, velocity-limited sources: $r=0.44$, p-value $\ll 0.001$). This correlation is expected since the solenoidal fraction takes into account the turbulent modes of the entire extension of a cloud. Larger clouds contain both compressive star-forming regions but also large envelopes of more rarefied gas around them, and the gas motions in these envelopes contribute to the increase in the values of $R$ for these clouds. This behaviour is also mirrored in the slight negative correlation between field size and SFE (full sample: $r = -0.13$, p-value $\ll 0.001$, velocity-limited source: $r=-0.18$, p-value $\ll 0.001$). Isolating the  $82$ clouds with solenoidal fraction $< 0.1$ that populate the upper left corner of Fig.~\ref{sf}) reveals that this set includes both compact cores (50-500 voxels) and small clouds (1500-3500 voxels). These clouds with associated luminosity from YSOs have a higher fraction of their gas in compressive turbulent modes. Their average velocity dispersion is 0.80 km\,s$^{-1}$. The distribution of the size of these clouds is shown in Fig.~\ref{upper_left}.

An evaluation of the effects of field-size correlation on the solenoidal-fraction-SFE relation through partial correlation analysis shows that the relation with field size does not account for the negative correlation between the solenoidal fraction and the SFE (a partial-Spearman test with field size as the co-dependent variable gives for the full sample: Spearman $r = -0.31$, p-value $\ll 0.001$ and for the velocity-limited sources: $r=-0.33$ and p-value $\ll 0.001$).

The correlation between the size and $R$ is lost when distance-dependent physical measures of size are considered. An example is the equivalent radius defined in \cite{Rigby2019}. This conclusion also holds when the volume of the cloud is expressed as number of voxels. 
Selecting a sub-sample (381) of sources with large field size ($> 65$ pixels), we recover a negative $L/M$-$R$ correlation similar to relation of fig.~ \ref{sf} with Spearman $r=-0.22$ and p-value $\ll 0.001$.

A prominent feature of the plot in Fig.~\ref{sf} is the scatter that characterises the relation between SFE and the solenoidal fraction. The scatter appears small at low $R$, increasing at higher values.  

We now examine whether this corresponds to a real change in the $L/M$ distribution with solenoidal fraction, or is due to larger sample sizes revealing the wings of the distribution, and whether it is due to measurement uncertainty or additional physical effects on $L/M$ and, hence, the SFE. 

Fig.~\ref{bolo} shows the solenoidal fraction-SFE scatter plot centred around its weighted linear fit (blue solid line in Fig.~\ref{sf}). The weights of the fit correspond to the standard deviations of the distributions of values of SFE obtained after binning the solenoidal fraction. 

The adjusted scatter plot in Fig.~\ref{bolo} displays a sharp increase of scatter in the SFE at $R \sim 10^{-1.3}$. To check that the distribution of $\log(L/M)$ at  $\log(R) <-1.3$ is statistically consistent with the distribution at $\log(R) > -1.3$, a Kolmogorov-Smirnov test is performed comparing the two distributions. As above, the null hypothesis is that the two samples are drawn from the same underlying population. 
With the Kolmogorov-Smirnov statistic $k = 0.37$ and p-value $= 0.022$, the null hypothesis cannot be rejected and the $\log$(SFE) distribution must be considered statistically consistent over the whole $\log(R)$ range, i.e., the scatter is not a function of $R$. 

The $L/M$ ratio is independent of distance, so the uncertainty associated with it equals the quadrature sum of the uncertainties in the flux and the column density. 

The latter is estimated to be $\sim$20\,{color{red}per}\,cent \citep{Rigby2019}.

The bolometric flux of a Hi-GAL source is evaluated using trapezium-rule integration over the five {\em Herschel} photometric bands.

The errors in the bolometric fluxes are therefore the quadrature sum of the uncertainties in each band.
The errors in the bolometric fluxes within a SCIMES cloud are then summed in quadrature to obtain the uncertainty associated with the whole cloud. This calculation yields an average error in the bolometric flux of $\sim$7\,per\,cent\footnote{Notice that the error in the bolometric flux is derived through the quadrature sum of the error at the five Hi-GAL wavelengths. Using a small number of wavelengths to estimate the error over the entire spectrum produces a lower value of the error. Thus one could say that the value from the Hi-GAL wavebands is a lower bound of the error in the bolometric flux.}.  The average relative measurement uncertainty in $L/M$ is, therefore, $\sim$21\,per\,cent. Since this is much smaller than the scatter in $L/M$, we must assume that there are one or more physical mechanisms other than the solenoidal fraction affecting the SFE within individual molecular clouds.  In the next section, we explore the most likely of these, i.e., the evolution of the embedded YSOs.

\subsection{Temperature and scatter}\label{temperature_and_scatter}

\begin{figure}
	\includegraphics[width=\columnwidth]{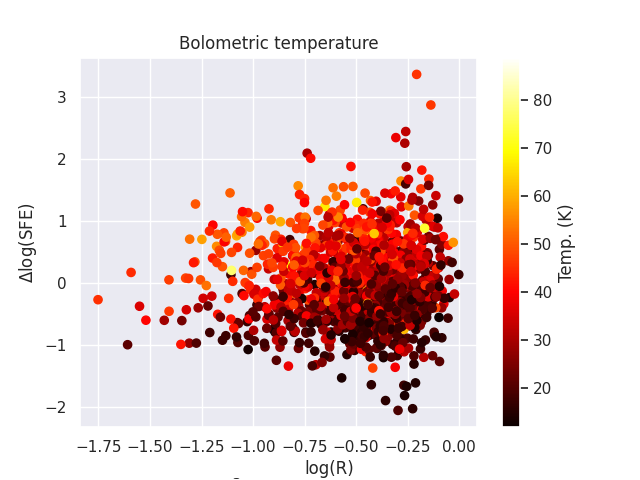}	
	\caption{Adjusted scatter plot of the SFE and solenoidal fraction for the full sample. The plot is centred around the weighted linear model shown in Fig.~\ref{sf}. Colour coding corresponds to the Hi-GAL bolometric temperature associated with each source. Luminosities and masses are given in units of $L_\odot$ and $M_\odot$, respectively.
	}
	\label{bolo} 
\end{figure}

In the case of dense clumps, where both $L$ and $M$ are derived from the same continuum data, the SFE measure $L/M$ is closely related to the bolometric temperature of the source \citep{Urquhart2018} and is, in fact, a temperature parameter, since $M = L \times f(T)$. In such circumstances, $L/M$ can be interpreted as a tracer of evolution for individual sources. Although we have here measured and defined $M$ and $L$ independently, we now consider the effect of temperature-traced evolution on the scatter in the data, using the Hi-GAL bolometric temperatures (colour-coded in Fig.~\ref{bolo})
of the embedded continuum sources.

The bolometric temperature is defined from the flux density $F_\nu$ \citep{Myers1993} as 

\begin{equation}
    T_\mathrm{bol} = 1.25\times 10^{-11} K \times
    \frac{\int^\infty_0 \nu F_\nu d\nu}{\int^\infty_0 F_\nu d\nu}.
\end{equation}

The temperature associated with each SCIMES cloud corresponds to the average
$T_\mathrm{bol}$ 
of the Hi-GAL sources it contains. In general, typical bolometric temperatures found in Hi-GAL clumps range
from $\sim 10$\,K (pre-stellar sources) to $\sim 80$\,K. 

\begin{figure*}
\includegraphics[width=\textwidth]{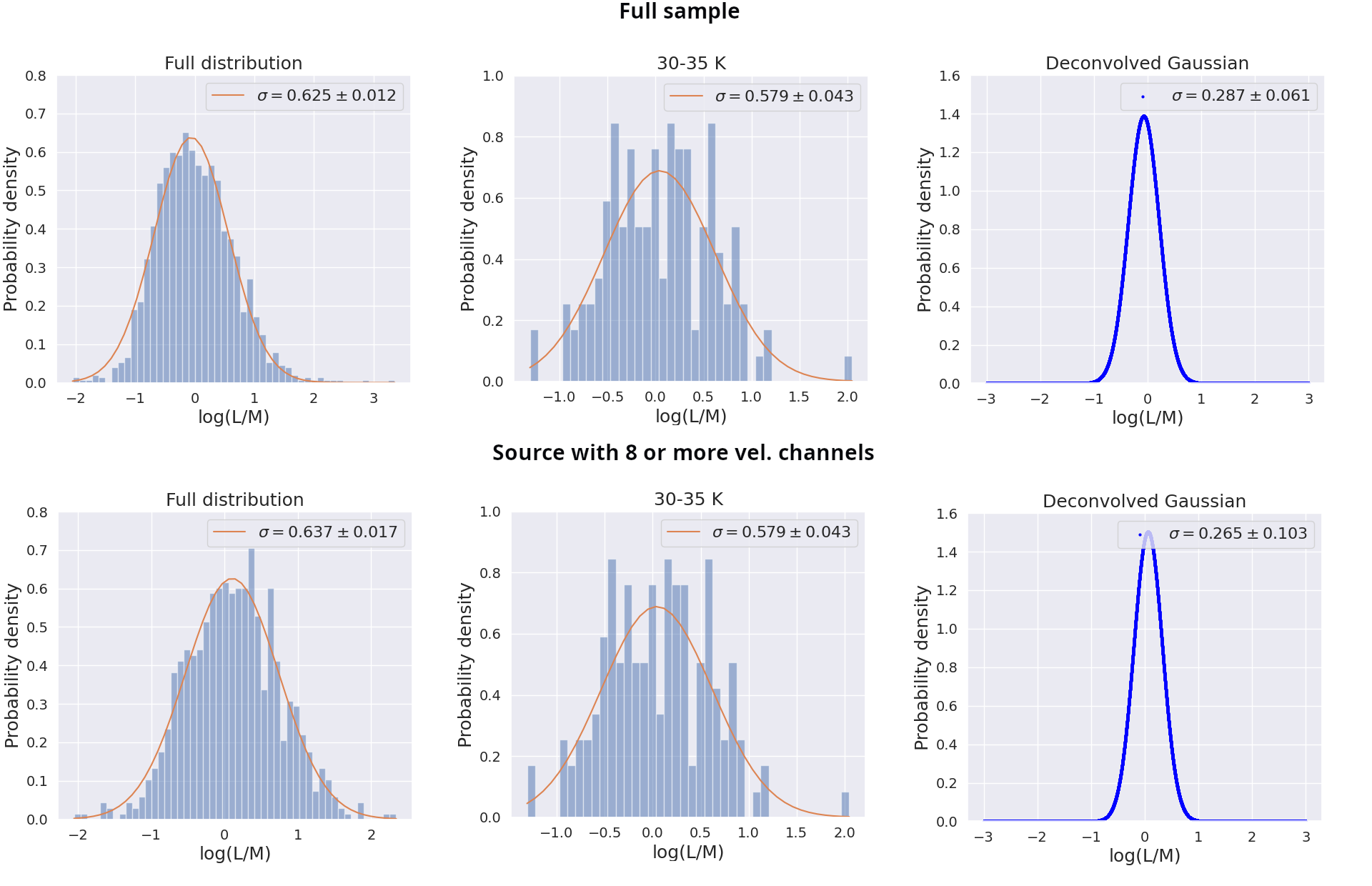}	
\caption{Deconvolutions of the full $L/M$ distributions by the Gaussians approximating the distributions of SFE in the $30-35$ K bin. The top and bottom panels refer to the full sample and the sources with 8 or more velocity channels respectively. Means and standard deviations are given for the fitted Gaussians. 
}

\label{decon} 
\end{figure*}

To quantify and filter out the scatter in $L/M$ that may be due to temperature and, hence, evolution variations, 
we select clouds in a narrow mean temperature range between 30 and 35\,K, in which the distribution of $L/M$ approximates a normal distribution. For the full sample, the mean $\log{L/M}$ in this range is $\mu = -0.074\pm 0.039$ with standard deviation $\sigma = 0.287\pm 0.062$ and for the velocity-limited sample, $\mu = 0.060\pm 0.057$ and $\sigma = 0.265\pm 0.103$  (see Fig.~\ref{decon}).  We take these subsets to be essentially free of evolution effects.

Estimating the variation coefficient ($c_v = \sigma /\mu$) of these `evolution-free' distributions
and converting it back to the linear scale gives $c_v = 2.03$ (the value of the standard deviation is 203\,per\,cent of the value of the mean) for the full sample and $2.21$ for the velocity-limited sources.
These values are still an order of magnitude larger than the measurement errors, suggesting that the residual scatter around the SFE-$R$ relation in Fig.\,\ref{sf} is the result of physical factors other than the solenoidal fraction that are involved in determining the SFE within clouds.


Fig.~\ref{decon} illustrates the deconvolution between the `evolution-free' Gaussian distributions described above and the full distribution containing the original scatter (Fig.\,\ref{bolo}). The result encodes the magnitude of the variation in $L/M$ arising from evolutionary effects.

There is no obvious correlation between Hi-GAL $T_{\rm bol}$ values in the present data and our independent CO masses, suggesting that the column density does not evolve significantly during the star-formation process. \cite{Urquhart2018} tested the corresponding correlation for the ATLASGAL clump sample, finding that the continuum-traced column density decreases as the clump evolves; however, they noted that the weak correlation found may arise from an observational bias: the reduced sensitivity to lower column densities. This therefore justifies the assumption made above that the cloud mass traced by \ce{^{13}CO} (3--2) does not change significantly over the IR-bright period of star formation traced by Hi-GAL data.

\section{Discussion and conclusions}
\label{conclusions}

Molecular clouds form through the condensation of the lower-density, atomic ISM gas, thus inheriting its turbulent and shear-driven motions \citep{Meidt2018, Meidt2019, Kruijssen2019c}. Galactic dynamics can thus stabilise clouds \citep{Meidt2013} or compress them promoting star formation \citep{Jeffreson2018}. In this framework, the relatively high star-formation efficiency (SFE) observed in disc clouds is linked to the prevalence of compressive (curl-free) turbulent modes. In contrast, the low SFE that characterises clouds in the Central Molecular Zone (CMZ) is related to the shear-driven solenoidal (divergence-free)
component. 
In this article, we performed a study of the turbulent modes in a large sample of Galactic-plane clouds from the \ce{^{13}CO}/\ce{C^{18}O} ($J=3\rightarrow 2$) Heterodyne Inner Milky Way Plane Survey (2266 sources), applying the method devised by \cite{Brunt2010} and \cite{Brunt2014}. 
This approach shows that that the majority of clouds are potentially star forming, having $R < 2/3$. Our analysis produced two main results:

\begin{itemize}

   \item there is a negative correlation between star-formation efficiency, as measured by the ratio of infrared luminosity to CO-traced cloud mass, and the relative power in the solenoidal modes of turbulence in the CO-traced gas, consistent with the hypothesis that solenoidal modes prevent or slow down the collapse of dense cores (Fig.~\ref{sf});
   
    \item the relative power in the solenoidal fraction appears to be highest in the inner Galaxy ($< 4$ kpc from the centre), declining in a shallow gradient with increasing Galactocentric distance. If confirmed by the analysis of a sample at lower longitudes, this result would be consistent with the disc becoming stable against gravitational collapse and the star formation rate being suppressed by the influence of the rotation of the Galactic bar, and/or with increased rotational shear at smaller radii;

\end{itemize}

These findings agree with the variation of SFE with the Galactic environment measured using both
the numbers of HII regions per unit molecular gas mass and the dense gas mass fraction (DGMF).
The DGMF peaks at radii around 3–4 kpc and then declines in the inner zone \citep{Eden2012, Eden2013}, where star formation is suppressed 
for the life of the bar in external barred-spiral galaxies \citep{James2009, James2016, Spinoso2017}.

Outside the Inner Galaxy, the solenoidal fraction 
declines monatonically in a shallow gradient, with no signal present at the spiral-arm radii. This latter result is in agreement with previous studies that found no significant arm-associated signal in the fraction of compact star-forming sources \citep{Ragan2016, Ragan2018}. 

This picture challenges the idea that spiral arms are direct triggers of star formation and considers them as mere producers of source crowding \citep{Moore2012, Ragan2016}. The increased star formation density observed in the spiral arms may be a consequence of their function as organising features that affect the ISM by delaying and crowding the gas that traverses them \citep{Dobbs2011}. Shear may represent a radial factor influencing this behaviour. \cite{Dib2012} calculated shear from a model rotation curve. Shear also appears to decline with radius and is a good candidate for much cloud formation via Kelvin-Helmholz type instabilities.

Shear (and tidal forces) may also introduce a bias in the calculated solenoidal fractions, disrupting the assumption of isotropy by inducing a preferred direction in the velocity field. The impact of this is likely to be strongest at short Galactocentric radii. 
A full investigation of the degree of loss of isotropy with Galactocentric radius would thus be required to assess the reliability of the solenoidal fraction values of all clouds in the sample. To some extent, the large sample of clouds analysed here should account for any effects of preferred viewing angle. However, an independent study of relatively nearby star-forming regions \citep{Zhou2022}, based on extension of the astrometric motions of YSOs to their surrounding gas, suggests isotropic turbulence, independent of both viewing angle and location. This gives some confidence that the assumption of isotropy is robust, but the relation between shear and solenoidal modes will be addressed in future work.

A prominent feature of the SFE-solenoidal fraction relation shown in Fig.~\ref{sf} is the scatter in the data, which is much larger than the measurement errors. Section \ref{SFE2} shows that the scatter remains an order of magnitude larger than the errors in a relatively evolution-free subset.
This remaining scatter indicates the presence of additional physical factors determining the value of $L/M$ and, by implication, the SFE in individual clouds.

Although compressive turbulence remains one of the driving agents of star formation in this framework, star-forming regions can be affected by several factors that slow down their collapse. In addition to delays induced by thermal pressure gradients at early stages of collapse, magnetic fields may play a role, even if the clouds are magnetically supercritical, i.e., the magnetic energy is less than the binding energy \citep{Inoue2009, Vazquez2011, Girichidis2018}. Galactic differential rotation through shear and Coriolis forces may be significant  \citep{Dobbs2014, Meidt2020} and the non-spherical (planar or filamentary) shapes of clouds \citep{Toala2012, Pon2012} contributes to moderating collapse times. 
For clustered star formation, numerical simulations show that stellar feedback such as protostellar jets, outflows, and stellar winds can inject supersonic turbulence in molecular clumps \citep{Nakamura2007, Offner2015}, and the clumps can be kept near virial equilibrium for several dynamical timescales. Al these factors may influence the observational snapshot of the SFE determined by the infrared-bright stage of the process.

We find that clouds with low values of solenoidal fraction and high SFE are usually spatially smaller, suggesting a lack of large envelopes of gas (see Fig.~\ref{upper_left}).  However, the value of the solenoidal fraction assigned to each cloud accounts for the overall modes of the gas it contains, with substructure contributing over all spatial frequencies. Thus, the same value of $R$ can be attained through different configurations of molecular gas, i.e., different cloud sizes, velocity distributions, densities, amount of molecular gas, number, and size of star-forming cores, and stellar feedback mechanisms.

\section*{Acknowledgements}

The authors would like to thank Jan Orkisz for helpful discussions on the treatment of noise power spectra. We thank Ana Duarte-Cabral and Dario Colombo for their help with SCIMES Python package. The Starlink software \citep{starlink} is currently supported by the East Asian Observatory.
This research made use of SCIMES, a Python package to find relevant structures into dendrograms of molecular gas emission using the spectral clustering approach \citep{scimes}. SCIMES is an astropy affiliated-package \citep{astropy:2013, astropy:2018}.

\section*{Data availability}

The data used for this paper are available from the archives of the
CHIMPS \citep{Rigby2019}, ATLASGAL \citep{Urquhart2018} and Hi-GAL \citep{Elia2017}.
The SCIMES catalogue used for the calculation of the solenoidal fraction and star formation efficiency is available to download from the CANFAR archive\footnote{\url{https://www.canfar.net/citation/landing?doi=22.0019}}.



\bibliographystyle{mnras}
\bibliography{example} 




\appendix

\section{The field size}\label{the_field_size}

\begin{figure}
	\includegraphics[width=\columnwidth]{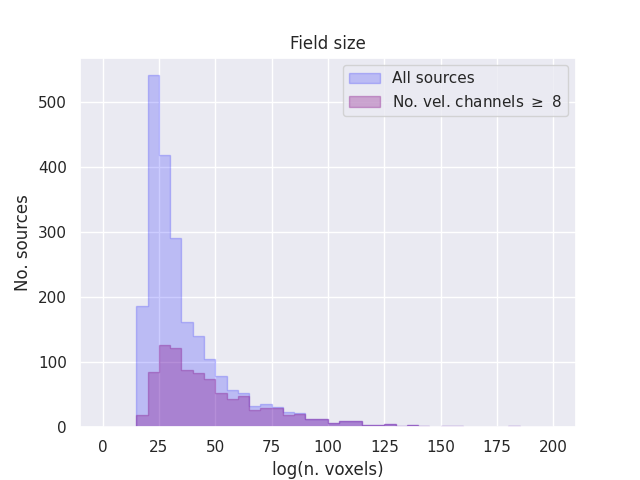} 
	\caption[Distribution of field sizes for calculation of the solenoidal fraction]{Distribution of field sizes including the 5-voxel padding for the sample of CHIMPS clouds used in calculation of the solenoidal fraction.
	}
	\label{fields} 
\end{figure}

The field sizes associated with the full sample of clouds considered in the analysis presented above range from 18 to 652 pixels in side (see Fig.~\ref{fields}). Evaluating the Fourier transforms on fields of small sizes is not likely to yield useful information on the state of the turbulence in the corresponding clouds, and so the validity of the results presented above was tested on a series of sub-samples with fields of decreasing size. This test showed that the results (distribution of solenoidal fraction with Galactocentric distance, the negative correlation between SFE and the solenoidal fraction, the scatter in solenoidal fraction-SFE plots) still hold when a sample of clouds with a field larger or equal to 85 pixels is considered. The sample size in this threshold case is reduced to less than 120 clouds. Above this threshold, the size of the sample is reduced drastically which invalidates the outcome of the analysis presented in subsection \ref{the_solenoidal_fraction} and \ref{SFE2}.
 
\bsp	
\label{lastpage}
\end{document}